\documentclass[sigconf]{acmart}

\usepackage{booktabs} 
\usepackage{comment}
\usepackage[font={footnotesize}]{caption}
\usepackage{float}

\setcopyright{none}
\acmConference{WSDM 2019 } {February XXX, 2019}{Melbourne, Australia.}  

\begin{document}
\title{EXS: Explainable Search Using Local Model Agnostic Interpretability}

 \author{Jaspreet Singh}
 \affiliation{%
   \institution{L3S Research Center, Hannover, Germany}
 }
 \email{singh@l3s.de}

 \author{Avishek Anand}
 \affiliation{%
   \institution{L3S Research Center, Hannover, Germany}
 }
 \email{anand@l3s.de}

\renewcommand{\shortauthors}{Singh and Anand}

\begin{abstract}
Retrieval models in information retrieval are used to rank documents for typically under-specified queries. Today machine learning is used to learn retrieval models from click logs and/or relevance judgments that maximizes an objective correlated with user satisfaction. As these models become increasingly powerful and sophisticated, they also become harder to understand. Consequently, it is hard for to identify artifacts in training, data specific biases and intents from a complex trained model like neural rankers even if trained purely on text features.
EXS is a search system designed specifically to provide its users with insight into the following questions: ``What is the intent of the query according to the ranker?'', ``Why is this document ranked higher than another?'' and ``Why is this document relevant to the query?''. EXS uses a version of a popular posthoc explanation method for classifiers -- LIME, adapted specifically to answer these questions. We show how such a system can effectively help a user understand the results of neural rankers and highlight areas of improvement.

\end{abstract}

%
%



\maketitle
\section{Introduction}


Neural networks have been widely successful in an array of learning tasks including vision, speech and NLP. Recently, the IR community has also seen successful application of deep learning models in document representation~\cite{mitra2017neural,chaidaroon2017variational}, query modeling~\cite{zamani2016queryLM,zamani2016estimating,diaz2016query} and design of ranking models~\cite{guo2016deep,desm16,mitra2016dual,wang2017irgan}. It is safe to assume that, as deep learning models continue to develop further, strides will be made towards making end-to-end neural information retrieval systems more viable. This will lead to increased deployment across several companies and institutions albeit as functional blackboxes due to a lack of transparency.  

In the context of machine learning, interpretability can be defined as ``the ability to explain or to present in understandable terms to a human"~\cite{doshi2017towards}. Interpretability is often deemed critical to enable effective real-world deployment of intelligent systems, albeit highly context dependent~\cite{weller2017challenges}. For a researcher or developer: to understand how their system/model is working, aiming to debug or improve it. For an  end user: to provide a sense for what the system is doing and why, to enable prediction of what it might do in unforeseen circumstances and build trust in the technology. Additionally, to provide an expert (perhaps a regulator) the ability to audit a prediction or decision trail in detail and establish legal liability, particularly if something goes wrong, e.g. explicit content for innocuous queries (for children) or expose biases that may be hard to spot with quantitative measures. This need has recently led to new tools and approaches for extracting explanations from black box models in tasks like image captioning, text classification and machine translation. However little work has been to done on building such tools to explain the output of ranking models.

Interpretability for query driven rankings is different from other prediction tasks like classification in the following aspects: the first major difference is that for classifiers we want to know why an instance was assigned a particular label whereas for rankings we want to know why a document is considered relevant to an arbitrary query. Secondly, for rankings one has to reason between pairs of documents -- explaining why one document was ranked above another. Lastly, since retrieval models encode a query intent which drives the output ranking, we should be able to explain what the model understands when a certain query is issued. This helps the user better diagnose the intent of the learnt model and isolate errors due to biases in the training data, artifacts of improper modeling and training etc. For example, why does the query 'jaguar' result in only animal related documents at the top.

\begin{figure*}[t!] 
\centering
\includegraphics[width=0.8\textwidth]{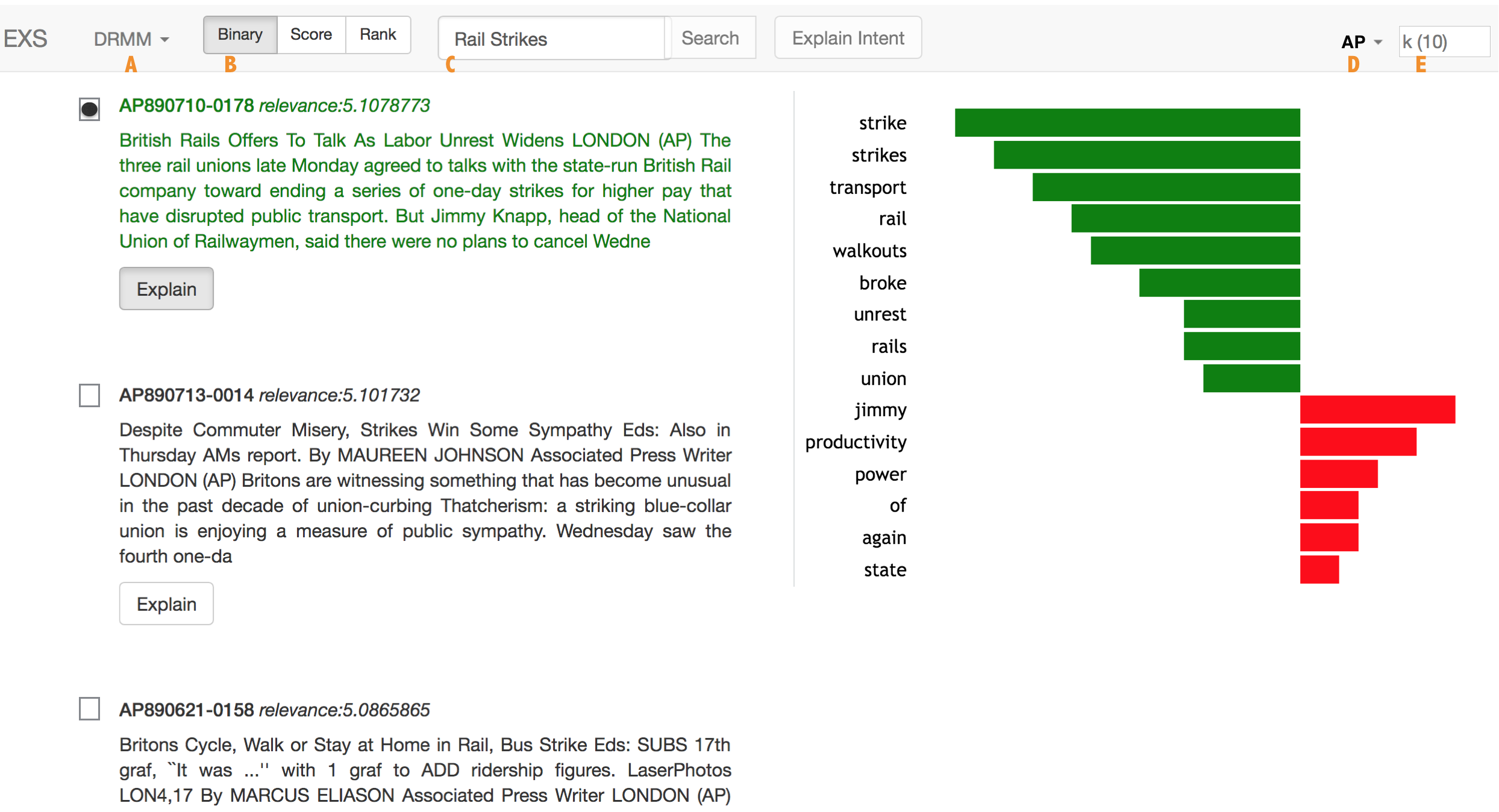}
\caption{The EXS User Interface. The top bar of the application houses the retrieval model selector (A), Score-to-Probability converter for LIME (B), search box (C) and the Explain Intent button in that order. To the right is the rank depth input box (E) and the corpus selector (D). The left pane shows the search results for the query 'Rail Strikes' according to DRMM. The right pane shows the output of clicking on the 'Explain' button corresponding to the top result. The bar chart on the right shows the words in the document that make it relevant and irrelevant according to DRMM. The green bar indicates the strength of a word for the relevant class and red for the irrelevant class. EXS can be found at \href{http://bit.ly/exs-search}{http://bit.ly/exs-search}}
\label{fig:ui} 
\end{figure*}

In this paper we describe EXS -- our EXplainable Search system designed to help end users better understand text based neural retrieval models. EXS' primary goal is to aid users in answering the following ranking-related questions: 
\begin{itemize}
    \item ``Why is this document relevant to the query?''
    \item ``Why is this document ranked higher than the other?''
    \item ``What is the intent of the query according to the ranker?''
\end{itemize}
Figure~\ref{fig:ui} illustrates the user interface of EXS. 

The remainder of this paper is organized as follows: first we highlight the broad range of interpretability techniques in ML in Section~\ref{sec:int}. Next we detail LIME~\cite{ribeiro2016model} (a pos-thoc model agnostic interpretability technique for classifiers) and our subsequent modification to deal with pointwise rankers in Section~\ref{sec:lime}. Then in Section~\ref{sec:implementation} we outline the architecture and implementation details of EXS. Furthermore we detail our choices when visualizing explanations to answer the 3 questions posed earlier. Finally in Section~\ref{sec:qualitative} we show anecdotal evidence of using EXS with a neural ranking model -- DRMM~\cite{guo2016deep}, to understand the ranking produced for a given query. EXS can be found at \url{http://bit.ly/exs-search}{http://bit.ly/exs-search}.

\section{Interpretability in Machine Learning} 
\label{sec:int}
Approaches to producing explanations can be categorized into two broad classes : \emph{model introspective} and \emph{model agnostic}. Model introspection refers to  ``interpretable'' models, such as decision trees, rules~\cite{rulesletham2015interpretable}, additive models~\cite{caruana2015intelligibletrees} and attention-based networks~\cite{captioningxu2015showattention}. Instead of supporting models that are functionally black-boxes, such as an arbitrary neural network, these approaches use models in which there is the possibility of meaningfully inspecting model components directly -- e.g. a path in a decision tree, a single rule, or the weight of a specific feature in a linear model.

Model agnostic approaches on the other hand like~\cite{ribeiro2016should,influencefunctionskoh2017understanding,ribeiro2018anchors} extract post-hoc explanations by treating the original model as a black box either by learning from the output of the black box model or perturbing the inputs or both. Model agnostic interpretability is of two types : local and global. \emph{Global interpretability} is concerned with approximating the overall behavior of a black box model with a simpler model. \emph{Local interpretability} on the other hand refers to the explanations used to describe a single decision of the model. 

\textbf{Interpreting Ranking Models} Recently there has been work on global interpretability for ranking models~\cite{singh2018posthoc} but little on how to explain black box retrieval models rank documents for a given query, i.e, a single result list. Ranking models can be pointwise -- output a score, pairwise -- output pair preference or listwise -- output a permutation. To build EXS we adapted a popular local model agnostic interpretability method for classifiers -- LIME~\cite{ribeiro2016should}, to explain the output of a pointwise ranker given a query. We choose pointwise rankers specifically since most neural ranking models are trained to predict a score given a query and document as input.

\section{Local Model Agnostic Interpretability for Ranking models}
\label{sec:lime}
 

 
Local model agnostic interpretability techniques~\cite{ribeiro2016should,seqalvarez2017causal, ribeiro2018anchors} produce explanations for a single decision by linearly approximating the local behavior of a complex blackbox model. This approximation is attained by training a simple linear model in a new interpretable feature space. The training data for the simple model is generated by perturbing the instance to explain constrained by the locality and querying the blackbox for labels. 

Approaches in this space tend to differ in the following aspects: the type of blackbox model (sequence prediction, regression, classification), definition of locality and the interpretable feature space. The interpretable feature space for text based models is usually the space of all words present in the instance. The explanation then is a visualization of the simple model depending on the context and the model itself. 
 
LIME~\cite{ribeiro2016should} is an approach designed specifically to explain the output of a classifier. To briefly illustrate their approach, consider a trained binary classifier $\mathcal{B}$ and an instance document to classify $d$. Assume that $\mathcal{B}(d)$ is a probability distribution across the classes. The objective of LIME is to train a simple model $\mathcal{M}_{d}$ that minimizes $\mathcal{L}(\mathcal{B}, \mathcal{M}_d, \pi_{d})$ which is a measure of how far $\mathcal{M}_{d}$ is in approximating $\mathcal{B}$ in the locality defined by $\pi_{d}$. The loss is to be reduced is the dissonance between the simple $\mathcal{M}_{d}$'s predictions and the labels produced by $\mathcal{B}$ for all $d' \in \pi_d$. $\mathcal{M}_{d}$ is a simple linear SVM trained on a feature space of words. $d' \in \pi_d$ is generated by perturbing $d$. In LIME, $d'$ is created by removing random words from random positions in $d$. Obtaining a label (probability distribution across classes here) for a perturbed instance is straightforward for a classifier whereas for ranking models this is tricky.

\textbf{From classifiers to rankers and back.} Let $q$ be a query and $D_q$ be the set of documents retrieved from the index. A pointwise ranker $\mathcal{R}$ produces a list of top-k documents ${D^{k}}_{q}$ after scoring each document with $\mathcal{R}(q,d)$ and then sorting. Now we discuss how ranking can be cast as a classification problem where the blackbox ranker predicts the class distribution of an arbitrary query-document pair. This is \emph{key to training} $\mathcal{M}_{d}$ on perturbed documents $d'$ using LIME.

Let $X$ be a random variable indicating the possible outcomes -- relevant and irrelevant. In the following we show how to estimate $P(X=relevant|q,d',\mathcal{R})$. Note that $P(X=irrelevant|q,d',\mathcal{R}) = 1 - P(X=relevant)$.

\textbf{Top-k Binary} In this case we make the assumption that $P(X=relevant|q,d',\mathcal{R}) = 1$ if $\mathcal{R}(q,d')$ is greater than $\mathcal{R}(q,{d'}_{k})$ where ${d'}_{k} \in {D^{k}}_{q}$ is the k-th document in the list.

\textbf{Score based} In this case we compute $P(X=relevant|q,d',\mathcal{R})$ as:

\begin{equation}
1 - \frac{\mathcal{R}(q,{d}_{1}) - \mathcal{R}(q,{d'})}{\mathcal{R}(q,{d}_{1})}
\end{equation}

where ${d}_{1} \in {D^{k}}_{q}$ is the top ranked document in the list. If $\mathcal{R}(q,d') \geq \mathcal{R}(q,{d}_{1})$ then $P(X=relevant) = 1$.

\textbf{Rank based} Here we look at the rank $d'$ in ${D^{k}}_{q}$. If $\mathcal{R}(q,{d'}) \leq \mathcal{R}(q,{d}_{k})$ then $P(X=relevant) = 0$. Otherwise, $1 - \frac{rank(d')}{k}$ where $rank(d')$ is the rank of the perturbed $d'$ in ${D^{k}}_{q}$.

\textbf{The Explanation Model} The approaches just discussed allow us to effectively train an explanation model $\mathcal{M}_{d}$ using the LIME framework. Since the feature space for $\mathcal{M}_{d}$ is a set of words and the model is a linear SVM, intuitively the sign and magnitude of the coefficients in $\mathcal{M}_{d}$ indicate which words in $d$ are strong indicators of relevance. In the next section (\ref{sec:viz}) we describe how we use the trained $\mathcal{M}_{d}$ to create visual explanations for the user.



\section{EXS -- An Explainable Search System}
\label{sec:implementation}
EXS was created to help users in understanding trained retrieval models using the modified version of LIME described above. The user interface of EXS, shown in Figure~\ref{fig:ui}, is that of a standard search engine: search results are displayed as a simple title and snippet. We reserve space on the right of the interface to display the explanation. Additionally in the header we allow the user to select from a list of trained retrieval models and text corpora. There is also an option to choose between Top-k binary, score based and rank based conversion for generating $\mathcal{M}$. 


\subsection{Visualizing Explanations}
\label{sec:viz}
We use the same bar chart style visualization as~\cite{ribeiro2016should} for $\mathcal{M}$. How this bar chart is generated however is dependent on the question the user is asking.
\begin{itemize}
\item[I] \emph{``Why is this document (d) relevant to the query?''} Each search result is equipped with an explain button which triggers an explanation model $\mathcal{M}_{d}$ to be trained. Once trained $\mathcal{M}_{d}$ tells us which words are strong indicators for either class by the sign and magnitude of their coefficients. Words which have a negative coefficient here are indicators of the relevant class. The magnitude values are plotted as a bar chart with green representing the relevant class. This is an artifact of LIME not our proposed modification. The explanation is visualized on the right as shown.

\item[II] \emph{``Why is this document($d_A$) ranked higher than another($d_B$)?''} A user can select a pair of search results using the checkbox next to the title. Once a pair is selected and the user clicks on explain, $\mathcal{M}_{d}$ is trained similar to the case before except $k=rank(d_B)$ and subsequently $d_k = d_B$. $\mathcal{M}_{d}$ now tells us which words in $d_A$ are strong indicators for either class when compared to the threshold set by $d_B$. 

\item[III] \emph{``What is the intent of the query according to the ranker?''} We assume that the intent of a query as per the selected $\mathcal{R}$ is a set of words $I$ whose presence in a document indicates high relevance, i.e. ranked in the top-k. Let $m_d$ be the set of words and corresponding coefficients for $\mathcal{M}_d$. We compute $I$ by aggregating $m_d$ for all $d \in {D^{k}}_{q}$. We simply add the coefficients of each word $w \in m_d$ for all $m_d$ and then select the top words and coefficients to display to the user. The graph is plotted the same way as before using $\sum m_d$ instead of a single $m_d$. A user can trigger an intent explanation by clicking on the explain button next to the search button. 
\end{itemize}

\subsection{Architecture and Implementation Details}

EXS (pronounced excess) is a client side web application that interacts with a REST API to retrieve search results and generate explanations. We used Lucene to index and retrieve documents from the standard news collections in TREC -- AP, LATIMES, Robust04 and Financial Times. Next we implemented and trained several neural networks like DRMM~\cite{guo2016deep} and DESM~\cite{desm16} using TensorFlow. We intend to extend both to include other models and text corpora over time. 

To produce explanations we used the implementation of LIME found in~\footnote{https://github.com/marcotcr/lime} with the modifications mentioned in Section~\ref{sec:lime}. LIME has two primary parameters when training $\mathcal{M}_d$ -- number of perturbed samples and number of words in explanation. We set the first to 2000 and give the user control over the second.

\section{Working with EXS}
\label{sec:qualitative}

\begin{figure}
\centering
\includegraphics[width=0.28\textwidth]{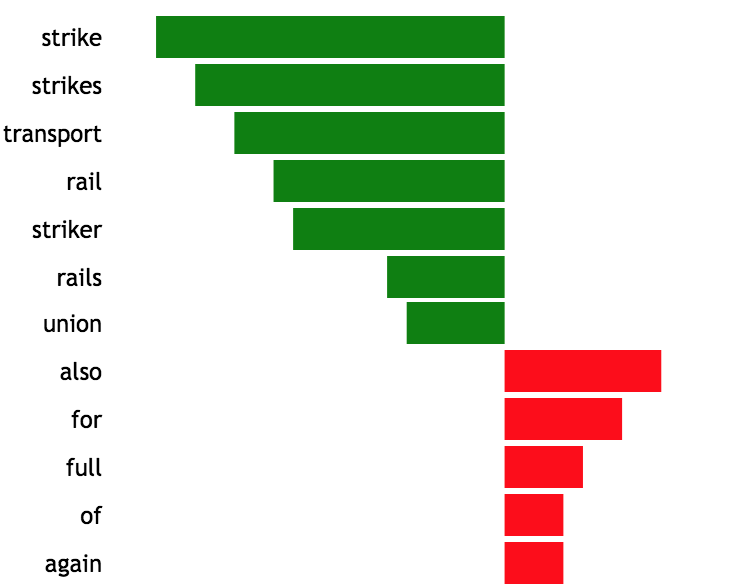}
\caption{Intent explanation for the query 'Rail Strikes' when using DRMM to rank documents from a news collection.}
\label{fig:intent}
\end{figure}

In this section we describe how users can use EXS to better understand decisions made by the Deep Relevance Matching Model~\cite{guo2016deep}. DRMM utilizes pertained word embeddings to first create query-document term interaction count histograms. This is fed as input to a feed forward neural network trained to predict a relevance score. Additionally it includes a gating mechanism to learn which parts of the query to pay attention to when scoring. We trained DRMM with the Robust04 TREC adhoc retrieval test collection. We used glove embeddings (300 dimensions) trained on the same to create the input histograms. 

Figure~\ref{fig:ui} shows the output of DRMM for the query 'Rail Strikes' for the Associated Press news collection. Clicking on the 'Explain Intent' button for this query results in Figure~\ref{fig:intent} being displayed on the right panel. We immediately notice that DRMM does capture the correct intent. We also find that words associated to the concept of a rail strike in particular like \texttt{transport} and \texttt{union} are strong indicators of the relevant class. Strong indicators of the irrelevant class are generic terms like \texttt{also} and \texttt{for}. 

\begin{figure}
\centering
\includegraphics[width=0.3\textwidth]{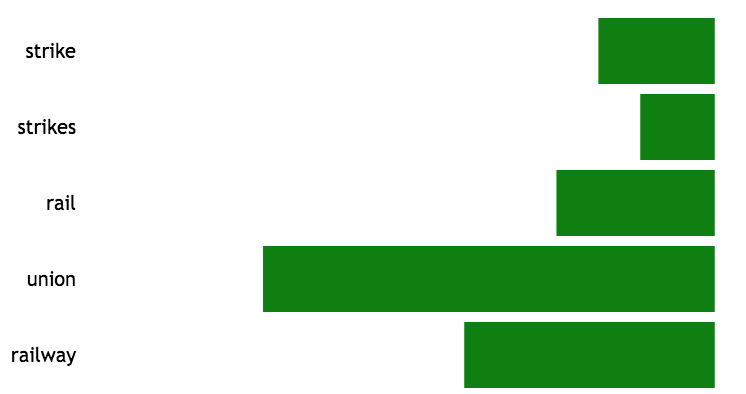}
\caption{Explanation for AP890710-0178 vs AP890713-0045 for the query 'Rail Strikes' when using DRMM}
\label{fig:pair}
\end{figure}

Clicking on the 'Explain' button under document AP890710-0178, EXS generates an explanation as to why AP890710-0178 is relevant to 'Rail Strikes' (shown in Figure~\ref{fig:ui}). The chart on the right is the visualization of $\mathcal{M}_d$ trained using the Top-k Binary method to convert the pointwise score output of DRMM to a probability distribution. Here we see words such as \texttt{unrest} and \texttt{walkouts} in AP890710-0178 that are particularly indicative of its' relevance as well as words like \texttt{jimmy} indicative of the contrary. This explanation indicates to the user that removing words like \texttt{jimmy} and \texttt{state} from AP890710-0178 can increase its' relevance whereas removing a word like \texttt{strike} can be highly detrimental.

The checkboxes next to each result allow for pair selection. Once AP890710-0178 and AP890713-0045 have been selected, clicking on 'Explain' in either result will generate an explanation as shown in Figure~\ref{fig:pair}. This explanation has only green bars since we're only interested in what makes AP890710-0178 more relevant. From the explanation it is clear that the words \texttt{railway} and \texttt{union} are the difference makers.

\section{Conclusion and Future Work}

In this paper we described our Explainable Search system EXS. EXS is designed to aid users in better understanding complex retrieval models like neural rankers. In this iteration of EXS we focus primarily on explaining pointwise neural rankers like DRMM. To allow for comparison by a variety of users and across a wide range retrieval models, EXS utilizes a posthoc model agnostic interpretability approach -- LIME, for generating explanations. We adapted LIME to deal with rankers since it was primarily created to explain the output of classifiers. To do so we had to devise approaches to convert the score output of a ranker (given a query and document as input) into a probability distribution across a relevant and irrelevant class. We leveraged the scores and rank positions of other documents in the top-k list to come up with 3 approaches: Top-k binary, score based and rank based. 

We then showed how the LIME explanations can be used to answer interpretability questions specific to ranking. We demonstrated 3 scenarios with the help of an anecdotal query and a neural ranker DRMM. EXS in its current state is a useful tool to diagnose and audit adhoc retrieval models. We also wish to address diversity based and multimodal retrieval models in the future. Another interesting area of future work is to investigate if integrating such diagnostics in medical and legal search applications (where interpretability is key) affects the user search behavior and how.


\bibliographystyle{ACM-Reference-Format}

\end{document}